\documentclass[prb,showpacs,twocolumn]{revtex4}%
\usepackage{amsfonts}
\usepackage{amsmath}
\usepackage{amssymb}
\usepackage{graphicx}
\usepackage{times}%
\begin{document}
\title{Datta-Das transistor: Significance of channel direction, size-dependence of
source contacts, and boundary effects}
\author{Ming-Hao Liu}
\affiliation{Department of Physics, National Taiwan University, Taipei 106, Taiwan}
\author{Ching-Ray Chang}
\affiliation{Department of Physics, National Taiwan University, Taipei 106, Taiwan}


\begin{abstract}
We analyze the spin expectation values for injected spin-polarized electrons
(spin vectors) in a [001]-grown Rashba-Dresselhaus two-dimensional electron
gas (2DEG). We generalize the calculation for point spin injection in
semi-infinite 2DEGs to finite-size spin injection in bounded 2DEGs. Using the
obtained spin vector formula, significance of the channel direction for the
Datta-Das transistor is illustrated. Numerical results indicate that the
influence due to the finite-size injection is moderate, while the channel
boundary reflection may bring unexpected changes. Both effects are concluded
to decrease when the spin-orbit coupling strength is strong. Hence [110] is a
robust channel direction and is therefore the best candidate for the design of
the Datta-Das transistor. \ \ \ \ 

\end{abstract}
\date{\today}
\maketitle

The Datta-Das spin-field-effect transistor (spin-FET),\cite{Datta-Das SFET}
stimulating plenty of theoretical and experimental works in semiconductor
spintronics,\cite{SM spintronics} has not yet been realized. Concluded
difficulties are basically:\cite{RMP: spintronics} (i) effective
controllability of the Rashba spin-orbit\cite{Rashba term} (SO) coupling
strength $\alpha$, (ii) long spin-relaxation time in two-dimensional electron
gas (2DEG) systems, (iii) uniformity of $\alpha$, and (iv) more efficient spin
injection rate. So far, the former two conditions have been basically
satisfied in experiments,\cite{Kikkawa JM 1998,Nitta J 1997} while the latter
two remain to be solved.

In the original proposal of the Datta-Das spin-FET, the structure inversion
asymmetry (the Rashba SO term) is required to dominate over the bulk inversion
asymmetry (the Dresselhaus SO term,\cite{Dresselhaus term} with coupling
strength $\beta$) therein. However, the coupling strengths of the Rashba and
Dresselhaus terms have been, in fact, found to be of the same order in certain
types of quantum wells.\cite{same order 1,same order 2} Therefore, the
influence due to the Dresselhaus term has become another issue in spintronics.
For example, \L usakowski \emph{et al.} had shown that the conductance of the
Datta-Das spin-FET depends significantly on the crystallographic direction of
the channel in the presence of the Dresselhaus term.\cite{Lusakowski A 2003} A
more complete work done by Winkler is the investigation of the spin-splitting
due to the effective magnetic field generated by the structure inversion
asymmetry and the bulk inversion asymmetry.\cite{Winkler's book,Winkler R
2004}

Recently, our previous work following Winkler even derived the analytical
formulae of the electron spin precession in the 2DEG with both the Rashba and
Dresselhaus terms involved.\cite{Liu MH 2005} The formulae obtained in Ref. 13
also implies the significance of the 2DEG channel direction, and is therefore
in correspondence with \L usakowski's result. However, the assumption of spin
injection via an ideal point contact and the neglect of boundary effects in
the 2DEG channel need be further investigated. In this paper, we mainly extend
our previous work\cite{Liu MH 2005} to include spin injection via finite-size
source contacts and to take the boundary effect into account. The former
consideration is found to provide an average effect and the change thus
induced is moderate, while the latter may bring drastic influences. Both
effects are concluded to be strong (weak) in weak (strong) SO-coupling
channels. In the case of $\alpha\beta>0$, electrons encounter the strongest
spin-splitting along the [110] direction,\cite{[110] spin splitting} which is
therefore concluded to be a robust channel direction as a good choice of the
Datta-Das spin-FET. Throughout this paper, we work within the single-particle
picture using a standard quantum mechanical approach, in particular, the
time-independent Schr\"{o}dinger picture, and assume zero temperature in the
clean limit.

Before considering the finite-size injection contact and the boundary effect
in the 2DEG channel, we first generalize the formulae obtained in Ref. 13,
which mainly describes the in-plane behavior of the electron spin, injected
from an inplane-magnetized ferromagnet into the 2DEG, via an ideal point
contact. Referring their results as $\left\langle \mathbf{S}\right\rangle
_{\mathbf{r}}^{\parallel}$ with the superscript denoting that the injected
spin is inplane-polarized while the subscript is for expectation done on
$\mathbf{r}=\left(  r,\phi\right)  $, we are now considering the more general
case, namely, spin injection with arbitrary polarization. The spinor
corresponding to the electron spin injected on $\mathbf{r}_{i}$ is therefore
given by\cite{Sakurai} $\left\vert \mathbf{s}\right\rangle _{\mathbf{r}_{i}%
}\doteq\left(
\begin{array}
[c]{cc}%
e^{-i\phi_{s}}\cos\left(  \theta_{s}/2\right)  , & +\sin\left(  \theta
_{s}/2\right)
\end{array}
\right)  ^{T}$, and we are thus seeking for the spin vector $\left\langle
\mathbf{S}\right\rangle _{\mathbf{r}}=\left(  \hbar/2\right)  \left\langle
\vec{\sigma}\right\rangle _{\mathbf{r}}$ with $\vec{\sigma}$ being the Pauli
matrices. Also, we present the calculation of $\left\langle \sigma
_{z}\right\rangle _{\mathbf{r}}$ to complete the description of the spatial
behavior of the spin vector. Using the same method introduced in Ref. 13, we
obtain{}, choosing $\mathbf{r}_{i}=\mathbf{0}$,%
\begin{equation}
\left\langle \vec{\sigma}\right\rangle _{\mathbf{r}}=\left(
\begin{array}
[c]{c}%
-\cos\theta_{s}\cos\varphi\sin\Delta\theta\left(  \mathbf{r}\right)
+\sin\theta_{s}\left\langle \sigma_{x}\right\rangle _{\mathbf{r}}^{\parallel
}\\
-\cos\theta_{s}\sin\varphi\sin\Delta\theta\left(  \mathbf{r}\right)
+\sin\theta_{s}\left\langle \sigma_{y}\right\rangle _{\mathbf{r}}^{\parallel
}\\
\cos\theta_{s}\cos\Delta\theta\left(  \mathbf{r}\right)  +\sin\theta
_{s}\left\langle \sigma_{z}\right\rangle _{\mathbf{r}}^{\parallel}%
\end{array}
\right)  \label{<S>3D}%
\end{equation}
with $\left\langle \sigma_{x}\right\rangle _{\mathbf{r}}^{\parallel}$ and
$\left\langle \sigma_{y}\right\rangle _{\mathbf{r}}^{\parallel}$ given by Eq.
(5) of Ref. 13, $\left\langle \sigma_{z}\right\rangle _{\mathbf{r}}%
^{\parallel}=\cos\left(  \varphi-\phi_{s}\right)  \sin\Delta\theta$,
$\varphi\equiv\arg[(\alpha\cos\phi+\beta\sin\phi)+i(\alpha\sin\phi+\beta
\cos\phi)]$, and $\Delta\theta\left(  \mathbf{r}\right)  =2m^{\ast}%
r\sqrt{\alpha^{2}+\beta^{2}+2\alpha\beta\sin\left(  2\phi\right)  }/\hbar^{2}$
with $m^{\ast}$ the electron effective mass, for the point spin injection case
in the absence of boundary effects. Clearly, Eq. (\ref{<S>3D}) recovers the
previous results in Ref. 13 when putting $\theta_{s}=\pi/2$.

Next we consider a spin-polarized source connected to the 2DEG channel, either
from the side or from the top, via a finite-size contact. Assume that each
electron is equally likely to be injected via all the possible injection
points, which may be everywhere of the contact except the positions close to
the atoms. Let us assume that the possible injection points locate on exactly
the center of each primitive unit cell of the contact crystal for the top
injection. In the case of side injection, the contact region becomes a line
and the injection points are reduced to the middle points of each neighboring
pair of atoms. Note that despite a displacement, the distribution of the
injection points are equivalent to the lattice points of the contact.
Labelling the positions of the injection points as $\mathbf{r}_{i}$, the state
ket describing the injected electron detected on $\mathbf{r}$ may be
superposed by $\left\vert \mathbf{s}\right\rangle _{\mathbf{r}}=\left(
1/_{\mathbf{r}}\left\langle \mathbf{s}|\mathbf{s}\right\rangle _{\mathbf{r}%
}\right)  \sum_{i}\left\vert \mathbf{s}\right\rangle _{\mathbf{r}%
_{i}\rightarrow\mathbf{r}}$ with $\left\vert \mathbf{s}\right\rangle
_{\mathbf{r}_{i}\rightarrow\mathbf{r}}$ given by\cite{Liu MH 2005} $\left\vert
\mathbf{s}\right\rangle _{\mathbf{r}_{i}\rightarrow\mathbf{r}}=\sum
_{\sigma=\pm1}\exp[-i\sigma\Delta\theta\left(  \mathbf{r}-\mathbf{r}%
_{i}\right)  /2]\left\langle \psi_{\sigma};\phi|\mathbf{s}\right\rangle
_{\mathbf{r}_{i}}\left\vert \psi_{\sigma};\phi\right\rangle $, where
$\left\vert \psi_{\sigma};\phi\right\rangle $ is the eigenspinor of the
Rashba-Dresselhaus system, $\phi$ being the angle of the wave vector. The
subscript $\mathbf{r}_{i}\rightarrow\mathbf{r}$ is to remind that the spin is
injected on $\mathbf{r}_{i}$ and detected on $\mathbf{r}$ after straight
evolution.\ Note that this formulation can also include the problem of
imperfect spin-polarized injection, i.e., to deal with a multi-domain
ferromagnetic source contact.

Now we deal with the channel boundary. The effect of the lateral confinement
in the channel was previously regarded as to provide a large energy gap
between two neighboring subbands to avoid intersubband mixing in the
transverse direction.\cite{Datta-Das SFET,Confinement,XDY} Contrary to this
suggested quasi-one-dimensional channel, we study a fully two-dimensional
channel and put emphasis on the electron wave property, i.e., both
longitudinal and transverse directions are not strongly quantized, so that the
spatial parts of the electron wave function in both directions are described
by plane waves (under the effective mass Hamiltonian). Based on this view
point, we assume that each spin-polarized electron injected on $\mathbf{r}%
_{i}$ may spatially evolve to the detection point $\mathbf{r}$ through not
only the straight but also the reflected paths. Treating the lateral
boundaries as hard walls, we write the corresponding state ket as%
\begin{equation}
\left\vert \mathbf{s}\right\rangle _{\mathbf{r}_{i}\rightarrow\mathbf{r}%
}^{\text{ref}}=\frac{1}{\text{ }_{\mathbf{r}_{i}\rightarrow\mathbf{r}%
}^{\text{ref}}\left\langle \mathbf{s}|\mathbf{s}\right\rangle _{\mathbf{r}%
_{i}\rightarrow\mathbf{r}}^{\text{ref}}}\sum_{n=0}^{\infty}\left\vert
\mathbf{s}\right\rangle _{\mathbf{r}_{i}\rightarrow\mathbf{r}}^{\left(
n\right)  }\text{,} \label{general state ket for reflection}%
\end{equation}
where $\left\vert \mathbf{s}\right\rangle _{\mathbf{r}_{i}\rightarrow
\mathbf{r}}^{\left(  n\right)  }$ is the spatially evolved state ket from
$\mathbf{r}_{i}$ to $\mathbf{r}$ after $n$ times of reflection by the channel
boundary. To avoid complicating the problem, we will pick terms up to $n=1$ in
the numerical results for the spin vectors under the influence of boundary
effects. Note that the source and drain contacts are assumed to be ohmic, and
hence we neglect the reflections in the longitudinal direction.

Note that to obtain the reflected waves is somewhat tricky. For example, the
$n=1$ term with $\mathbf{r}_{i}\rightarrow\mathbf{r}^{\prime}\rightarrow
\mathbf{r}$ path, where $\mathbf{r}^{\prime}$ is the position vector the
reflection occurs, can be obtained by $\left\vert \mathbf{s}\right\rangle
_{\mathbf{r}_{i}\rightarrow\mathbf{r}^{\prime}\rightarrow\mathbf{r}}%
=\sum_{\sigma}\exp[-i\sigma\Delta\theta(\mathbf{r}-\mathbf{r}^{\prime
})/2]\left\langle \psi_{\sigma};\phi_{\mathbf{r}-\mathbf{r}^{\prime}%
}|\mathbf{s}\right\rangle _{\mathbf{r}_{i}\rightarrow\mathbf{r}^{\prime}%
}\left\vert \psi_{\sigma};\phi_{\mathbf{r}-\mathbf{r}^{\prime}}\right\rangle $
with $\left\vert \mathbf{s}\right\rangle _{\mathbf{r}_{i}\rightarrow
\mathbf{r}^{\prime}}=\sum_{\sigma}\exp[-i\sigma\Delta\theta(\mathbf{r}%
^{\prime}-\mathbf{r}_{i})/2]\left\langle \psi_{\sigma};\phi_{\mathbf{r}%
^{\prime}-\mathbf{r}_{i}}|\mathbf{s}\right\rangle _{\mathbf{r}_{i}}\left\vert
\psi_{\sigma};\phi_{\mathbf{r}^{\prime}-\mathbf{r}_{i}}\right\rangle $,
$\phi_{\mathbf{r}}$ being the argument of the vector $\mathbf{r}$. When
considering finite-size injection, the total state ket characterizing the
injected electron is expressed as $\left\vert \mathbf{s}\right\rangle
_{\mathbf{r}}=\left(  1/_{\mathbf{r}}\left\langle \mathbf{s}|\mathbf{s}%
\right\rangle _{\mathbf{r}}\right)  \sum_{i}\left\vert \mathbf{s}\right\rangle
_{\mathbf{r}_{i}\rightarrow\mathbf{r}}^{\text{ref}}$ with $\left\vert
\mathbf{s}\right\rangle _{\mathbf{r}_{i}\rightarrow\mathbf{r}}^{\text{ref}}$
given by Eq. (\ref{general state ket for reflection}).

Now we present the calculated spin vectors inside the 2DEG channel with
certain cases of spin injection. We investigate InGaAs 2DEG channels, setting
the Rashba coupling parameter\cite{alpha = 3 eVnm} $\alpha=0.3%
\operatorname{eV}%
\operatorname{\text{\AA}}%
$ with electron effective mass $m^{\ast}=0.03m_{e}$, in the spin transistor
geometry, i.e., injected spin-polarization parallel to the channel direction.
The Dresselhaus coupling parameter is chosen as $\beta=0.09%
\operatorname{eV}%
\operatorname{\text{\AA}}%
$, which is deduced from $\beta\approx\gamma\left\langle k_{z}^{2}%
\right\rangle $ with a typical value for the coefficient\cite{same order
1,beta = 0.9 eVnm} $\gamma\approx25%
\operatorname{eV}%
\operatorname{\text{\AA}}%
^{3}$, assuming an infinite quantum well in $z$-direction with well width $50%
\operatorname{\text{\AA}}%
$.%
\begin{figure}
[ptb]
\begin{center}
\includegraphics[
height=2.3687in,
width=3.3978in
]%
{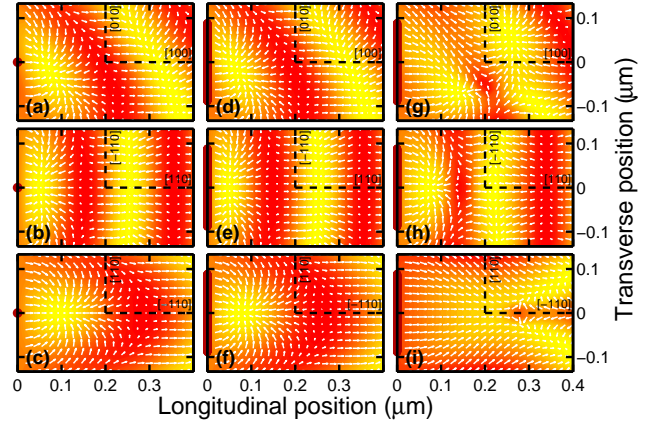}%
\caption{(Color online) Spin vectors in $0.4\operatorname{\mu m}%
\times0.267\operatorname{\mu m}$ InGaAs 2DEG channels using (a) -- (c) point
spin injection and (d) -- (i) finite-size spin injection, with the source
contacts (indicated by dark dots or dark thick lines in each panel) all
polarized parallel to the channel directions. Channel boundary effect is
considered in (g) -- (i). Color shading is determined by $\left\langle
S_{z}\right\rangle $ with red (dark) $\rightarrow$ negative and yellow
(bright) $\rightarrow$ positive.}%
\label{fig1}%
\end{center}
\end{figure}

We begin with the single-point spin injection case for different channel
directions without boundary effects. Three $0.4%
\operatorname{\mu m}%
\times0.267%
\operatorname{\mu m}%
$ 2DEG channels along [100], [110], and [1\={1}0] are examined, and the spin
is injected on the middle point of the left end. Using Eq. (\ref{<S>3D}), we
sketch the spin vectors inside the three channels in Figs. \ref{fig1}(a) --
(c). Different spin patterns shown in the three cases indicate that the
$z$-rotational symmetry is broken due to the presence of the Dresselhaus term.
Thus the influence due to the bulk inversion-asymmetry is clear, even in this
Rashba dominating 2DEG. As suggested in our previous work,\cite{Liu MH 2005}
channel directions should be chosen along [1$\pm$10] since the Rashba and
Dresselhaus terms generate a $k$-dependent effective magnetic field, which is
perpendicular to the electron propagation only along these two directions.
Such a uniqueness of these two axes we have just shown also agrees with
Averkiev's conclusion that [1$\pm$10] are the principle axes of the spin
relaxation rate tensor.\cite{Averkiev}

We now consider finite-size spin injection. The width of the contact is set
2/3 times the channel width, and perfect polarization of the source contact in
the spin transistor geometry is assumed. The spin vectors are plotted in Figs.
\ref{fig1}(d) -- (i), where the middle and right columns are in the absence
and in the presence of the boundary effect, respectively. Compared to the
single-point injection shown in Figs. \ref{fig1}(a) -- (c), the variation due
to the finite-size spin injection seems tiny in the case without the boundary
effect [Figs. \ref{fig1}(d) -- (f)], and the assumption of single-point
injection may thus work well in this case. When the boundary effect is taken
into account, the spin vectors are drastically changed [Figs. \ref{fig1}(g) --
(i)]. Comparing (d) -- (f) with (g) -- (i), respectively, one can roughly
conclude that the influence due to the boundary effect is stronger (weaker) in
weaker (stronger) SO-coupling channels, recalling that this spin-splitting is
strongest along [110] whereas that along [1\={1}0] is weakest when
$\alpha\beta>0$. In fact, this is also true for the influence due to the
finite-size injection, as will be clearer later.%
\begin{figure}
[ptb]
\begin{center}
\includegraphics[
height=1.2522in,
width=3.4229in
]%
{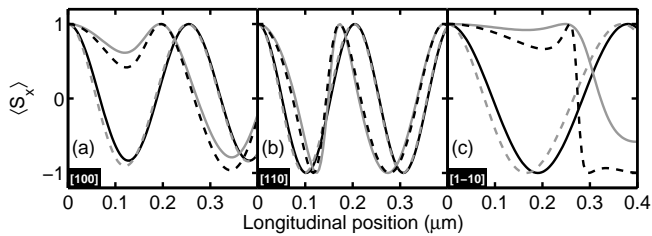}%
\caption{The $x$-component $\left\langle S_{x}\right\rangle $ in units of
$\hbar/2$ along the straight path in the middle of a (a) [100], (b) [110], and
(c) [1\={1}0] channel. Legend: black-solid and gray-dashed lines depict
single-point and finite-size injection cases, respectively, without the
boundary effect; gray-solid and black-dashed lines depict single-point and
finite-size injection cases, respectively, in the presence of the boundary
effect.}%
\label{fig2}%
\end{center}
\end{figure}

To specify the change due to the finite-size spin injection and the channel
boundary effect, we analyze $\left\langle S_{x}\right\rangle $ along the
straight paths in the middle of each channel, as shown in Fig. \ref{fig2}.
Without boudary effect, the effect of the finite-size injection merely reduces
the spin precession lengths near the source contacts, and the corresponding
reduction seems clearest and vaguest in the weakest ([1\={1}0]) and strongest
([110]) SO-coupling channels, respectively. This behavior becomes even clearer
when the boundary effect is present. Comparing black-dashed (finit-size
injection) with gray-solid lines (point injection), the difference seems,
again, clearest (vaguest) in the [1\={1}0] ([110]) channel. When focusing on
individually the effect of the boundary reflection, similar behavior is
observed. Taking the single-point injection cases (black- and gray-solid lines
in Fig. \ref{fig2}) for illustration, the change due to the boundary effect
appears again the most drastic in the case of [1\={1}0], where the spin
precession behavior is almost destroyed.

\begin{figure}
[ptb]
\begin{center}
\includegraphics[
height=1.3223in,
width=3.4229in
]%
{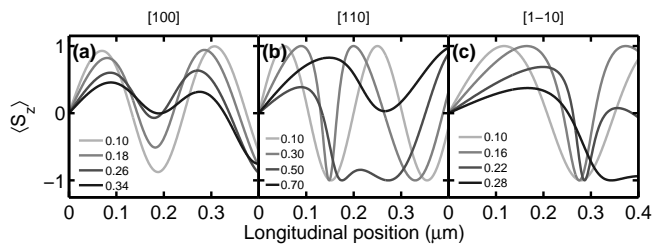}%
\caption{The $z$-component $\left\langle S_{z}\right\rangle $ in units of
$\hbar/2$ along the straight path in the middle of a (a) [100], (b) [110], and
(c) [1\={1}0] channel with different channel widths. The legends in each panel
label the corresponding channel widths in units of $\operatorname{\mu m}$.}%
\label{fig3}%
\end{center}
\end{figure}

Moreover, such interference effect may grow with the increase of the channel
width. Figure \ref{fig3} shows $\left\langle S_{z}\right\rangle $ under the
influence of the boudary effect and under the same conditions with Fig.
\ref{fig2}, except the varying channel widths. Clearly, we see that the
interference effect grows fastest with the increase of the channel width in
the [1\={1}0] channel, while the most slowly in the [110] channel, implying
its robustness characteristic. Also, the precession behavior is destroyed when
the channel width is wide enough. This conclusion that when one considers
wider (narrower) channels, the influence of the boundary effect becomes
stronger (weaker), also agrees with the previous suggestions of using
quasi-one-dimensional channel for enhancing the performance of the Datta-Das
transistor,\cite{Confinement} and also the slowdown of the D'yakonov-Perel'
spin relaxation rate in narrow channels.\cite{Malshukov}%

\begin{figure}
[b]
\begin{center}
\includegraphics[
height=2.1689in,
width=3.0113in
]%
{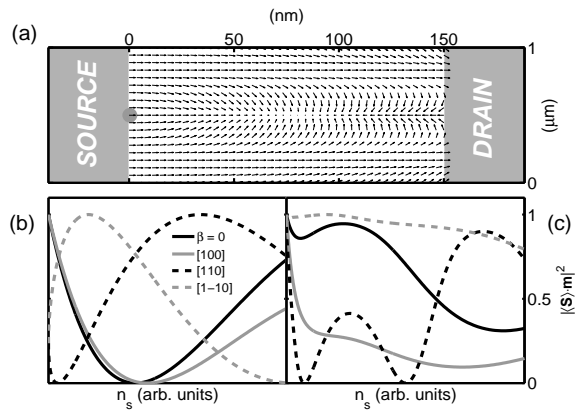}%
\caption{(a) Spin vectors inside a Rashba-type 2DEG channel of the spin-FET.
Electron spins are injected via the middle point of the source contact. (b)
and (c) plot $\left\vert \left\langle \mathbf{S}\right\rangle \cdot
\mathbf{m}\right\vert ^{2}$ as a function of the carrier density $n_{s}$ of
the 2DEG, determined on the middle point at the end of the channel in the
absence and in the presence of the boundary effect, respectively. The ranging
of the carrier density $n_{s}$ correspond to $\alpha=0-0.22\operatorname{eV}%
\operatorname{\text{\AA}}$.}%
\label{fig4}%
\end{center}
\end{figure}

We finally make a simple connection to the ballistic spin transport, solving
for the transmission problem in a ferromagnet-2DEG-ferromagnet double junction
structure, constructed by Matsuyama \textit{et al.,}\cite{Mastuyama} who
considered only the Rashba term in the 2DEG channel. As shown in Fig.
\ref{fig4}(a), we analyze the inplane components of the spin vectors in a $150%
\operatorname{nm}%
\times1.0%
\operatorname{\mu m}%
$ Rashba-type 2DEG (for consistency with their work), using the method of
single-point injection.%

From the time-independent Schr\"{o}dinger equation, one can obtain the
corresponding transmission probabilities, which are found to depend on (i) the
Fermi velocity mismatch between the ferromagnet and 2DEG regions and (ii) the
spinor overlap between the incoming and outgoing states. Of particular
interest is that the transmission probabilities (and, in fact, also the
reflection probabilities) are proportional to this spinor overlap. Put in
another way, the electron dislikes changing its spin direction when crossing
the boundary between the ferromagnet and 2DEG regions. In this sense, one can
clearly see why the oblique injections contribute an "undesired background" to
the transmission probability, and hence the conductance,\cite{Mastuyama} by
noting that the spin vectors at the right end of the channel in Fig.
\ref{fig4}(a) are nonuniform. Assuming that the drain contact is polarized
parallel to the channel direction, then only the spin vectors pointing to
right on the 2DEG-drain interface give positive contribution to the
conductance. Since only the normally injected spins in this Rashba channel
encounter the precession axis parallel to the 2DEG-drain interface, giving
rise to the maximum oscillating amplitude when varying the $\alpha$, the
oscillation behavior of the total conductance obtained by summing all the
transmission amplitudes from all the transmission modes (Landauer formula)
will eventually be averaged down.

Focusing on the normal injection [center path in the channel of Fig.
\ref{fig4}(a)], the spin vectors indeed precess upright down the way to the
drain, as the original design of the Datta-Das spin-FET.\cite{Datta-Das SFET}
Using the relation\cite{Mastuyama} $n_{s}\propto\alpha^{2}$, we plot
$\left\vert \left\langle \mathbf{S}\right\rangle \cdot\mathbf{m}\right\vert
^{2}$ ($\left\langle \mathbf{S}\right\rangle $ is determined at right end of
the channel, and $\mathbf{m}\ $is the unit vector of the channel direction),
which is responsible for the transmission probabilities $T_{\pm,\pm}$, as a
function of the carrier density $n_{s}$ in Figs. \ref{fig4}(b) and (c). In the
absence the boundary effect [Fig. \ref{fig4}(b)], the squared projection
$\left\vert \left\langle \mathbf{S}\right\rangle \cdot\mathbf{m}\right\vert
^{2}$ is isotropic and shows no dependence on the crystallographic direction,
when the Dresselhaus term is not involved. Despite the rapid oscillations
caused by the Fabry-Perot interference between the source-2DEG and 2DEG-drain
boundaries, we obtain a satisfactory curve [black-solid line in Fig.
\ref{fig4}(b)], in good agreement with Ref. 22 [see Figs. 10(c) -- (f)
therein]. We again stress on the importance of choosing the channel direction
for the Datta-Das transistor with $\beta\neq0$, by noting that the precession
behavior of the injected spin is sensitive to the channel direction [the other
three lines in Fig. \ref{fig4}(b)]. When the boundary effect is involved [Fig.
\ref{fig4}(c)], the spin precession behavior is totally changed, even the
robust [110] channel. This is because the channel width is too wide, allowing
a much severer influence caused by the boundary reflection, as we have
discussed previously.

In conclusion, we have calculated the spin vectors inside the 2DEG channel of
the Datta-Das transistor to demonstrate the significance of the channel
direction and to investigate the size-dependence of source contacts and the
channel boundary effects. The analytical spin vector formulae for the point
spin injection\cite{Liu MH 2005} are also generalized to arbitrary
polarization of spin injection cases. Numerical results have shown that the
influence due to the finite-size injection is moderate, while the channel
boundary reflection may bring unexpected changes.

We emphasize here the two-dimensional wave property of the electron in typical
InGaAs 2DEGs. From degenerate perturbation theory, one is led to the
criterion\cite{Datta-Das SFET,Confinement} $W\ll\hbar^{2}/\alpha m^{\ast}$,
within which the channel can be regarded as quasi-one-dimensional, when
assuming hard wall lateral confinement and considering only the Rashba term.
For the Rashba-Dresselhaus 2DEGs with, e.g., $\alpha=0.3%
\operatorname{eV}%
\operatorname{\text{\AA}}%
$, $\beta=0.09%
\operatorname{eV}%
\operatorname{\text{\AA}}%
$, and $m^{\ast}=0.03m_{e}$, the total coupling strength ranges from $0.21%
\operatorname{eV}%
\operatorname{\text{\AA}}%
$ ( the [1\={1}0] direction) to $0.39%
\operatorname{eV}%
\operatorname{\text{\AA}}%
$ (the [110] direction). The criterion for the quasi-one-dimensional channel
will require $W\ll65%
\operatorname{nm}%
$ ($121%
\operatorname{nm}%
$) for the [110] ([1\={1}0]) case. Therefore, typical InGaAs 2DEGs with
channel widths of the order of or larger than these lengths will require
two-dimensional description for the electron waves, and the possible boundary
effects are thus unavoidable.

Our results may be taken as a warning indicating another difficulty inherent
in the design of the Datta-Das spin-FET: the interference due to the channel
boundaries. However, we conclude that the [110] direction is shown to be
robust under the influence of finite-size spin injection and the boundary
reflection, for [001]-grown zincblend-based 2DEGs.$\allowbreak$

This work was supported by the Republic of China National Science Council
Grant No. 94-2112-M-002-004.

\end{document}